\documentclass[pdflatex,sn-mathphys-num]{sn-jnl}

\usepackage{graphicx}%
\usepackage{multirow}%
\usepackage{amsmath,amssymb,amsfonts}%
\usepackage{amsthm}
\usepackage[title]{appendix}%
\usepackage{xcolor}%
\usepackage{textcomp}%
\usepackage{subfigure}
\usepackage{manyfoot}%
\usepackage{booktabs}%
\usepackage{algorithm}%
\usepackage{algorithmicx}%
\usepackage{algpseudocode}%
\usepackage{listings}%
\usepackage{txfonts}
\usepackage{bm}
\usepackage{tabularx}

\usepackage{geometry}
\newcommand{\reply}[1]{{\color{black}{#1}}}

\usepackage{xcolor}

\usepackage[normalem]{ulem}
\usepackage{cancel}
\newif\ifrevision
\revisiontrue 

\ifrevision
  
  \newcommand{\del}[1]{\textcolor{red}{\sout{#1}}}

  \newcommand{\note}[1]{\textcolor{teal}{[Note: #1]}}
\else
  
  \newcommand{\del}[1]{}
  
  \newcommand{\note}[1]{}
\fi

\begin{document}
\title[Article Title]{Testing the strong equivalence principle with multimessenger binary neutron star mergers}

\author[1]{\fnm{Jie} \sur{Zhu}}
\equalcont{These authors contributed equally to this work.}
\author*[2,3]{\fnm{Hanlin} \sur{Song}}\email{hanlin@stu.pku.edu.cn}
\equalcont{These authors contributed equally to this work.}
\author*[4]{\fnm{Zhenwei} \sur{Lyu}}\email{zwlyu@dlut.edu.cn}
\author*[1]{\fnm{Hao} \sur{Li}}\email{haolee@cqu.edu.cn}
\author[3]{\fnm{Peixiang} \sur{Ji}}
\author[5,3]{\fnm{Jun-Chen} \sur{Wang}}
\author[3]{\fnm{Haobo} \sur{Yan}}
\author*[6,3]{\fnm{Bo-Qiang} \sur{Ma}}\email{mabq@pku.edu.cn}

\affil[1]{\centering \orgdiv{Department of Physics and Chongqing Key Laboratory for Strongly Coupled Physics},\\ \orgname{Chongqing University},
\orgaddress{\city{Chongqing}, \postcode{401331}, \country{China}}}

\affil[2]{\orgdiv{Center for Gravitational Wave Experiment, National Microgravity Laboratory, Institute of Mechanics}, \orgname{Chinese Academy of Sciences}, \city{Beijing}, \postcode{100190}, \country{China}}

\affil[3]{\orgdiv{School of Physics}, \orgname{Peking University},
\orgaddress{\city{Beijing}, \postcode{100871}, \country{China}}}

\affil[4]{\orgdiv{Leicester International Institute}, \orgname{Dalian University of Technology},
\orgaddress{\city{Panjin}, \postcode{124221}, \country{China}}}

\affil[5]{\orgdiv{\centering  Department of Physics and Jockey Club Institute for Advanced Study}, \\\orgname{The Hong Kong University of Science and Technology},
\orgaddress{\city{Hong Kong S.A.R.}, \country{China}}}

\affil[6]{\orgdiv{School of Physics}, \orgname{Zhengzhou University},
\orgaddress{\city{Zhengzhou}, \postcode{450001}, \country{China}}}

\abstract{
The constancy of the gravitational constant $G$ is a cornerstone of the strong equivalence principle and of general relativity, yet its possible temporal variation remains a key target in tests of fundamental physics. Gravitational-wave (GW) astronomy, especially when combined with electromagnetic observations, provides a powerful way to probe this principle in the strong-field and dynamical regime. 
In this work, we develop a GW waveform model with a slowly varying gravitational constant, incorporating its effects both on compact binary dynamics and GW propagation in an expanding universe. Applying this framework to the binary neutron star merger GW170817, together with independent electromagnetic constraints on the luminosity distance, sky localization and binary inclination from GRB 170817A, we perform a joint Bayesian analysis that disentangles varying-$G$ effects from astrophysical degeneracies. 
We find no evidence for a temporal variation of the gravitational constant, and constrain its fractional time derivative to $\dot{G}/G \in [-3.36 \times 10^{-9}, 5.34\times10^{-10}]~{\rm yr^{-1}}$, representing the most stringent bounds obtained to date from real GW observations. 
}

\maketitle

Einstein's theory of general relativity (GR) is one of the pillars of modern physics and has passed a wide range of experimental tests to date with remarkable precision, from the Solar System to astrophysical and cosmological scales~\cite{Ghez:2003qj, Psaltis:2008bb, Will:2014kxa, Freire:2024adf}. Among its foundational principles, the strong equivalence principle (SEP) plays a central role: it states that all local nongravitational and gravitational experiments, including those involving gravitational self-energy, are independent of the location, velocity, and composition of the freely falling reference frame. A direct consequence of the SEP is the constancy of Newton’s gravitational constant $G$. 
While the SEP has been stringently tested using lunar laser ranging, pulsar timing, the cosmic microwave background, and other observations~\cite{Williams:2004qba, Hofmann:2018myc, Kaspi:1994hp, Thorsett:1996fr, Deller:2008jx, Zhu:2015mdo, Zhu:2018etc, Wu:2009zb}, most existing constraints probe weak-field or quasi-static regimes. Testing the SEP in the strong-field, highly dynamical regime therefore remains an important open challenge.

The advent of gravitational-wave (GW) astronomy has opened a new window on gravity in precisely this regime. In particular, the binary neutron star (BNS) merger GW170817, detected by Advanced LIGO and Advanced Virgo on August 17, 2017~\cite{LIGOScientific:2017vwq}, marked the first observation of compact-object coalescence through both gravitational and electromagnetic channels. The detection of the associated short gamma-ray burst GRB 170817A just 1.7 seconds after the merger~\cite{LIGOScientific:2017zic}, followed by extensive multiwavelength follow-up observations~\cite{LIGOScientific:2017ync}, provided independent measurements of the
sky localization, luminosity distance, and orbital inclination of the source~\cite{DES:2017kbs, Cantiello:2018ffy, Mooley:2017enz, Mooley:2018qfh, Mooley:2022uqa}.
This combination of GW and electromagnetic information enables a genuine multimessenger test of gravity, breaking key parameter degeneracies inherent to GW-only analyses and allowing for precision probes of fundamental principles in the strong-field and dynamical regime.

{
Following the multimessenger observation of GW170817/GRB 170817A, extensive efforts have been devoted to probing modified gravitational-wave propagation. Belgacem \textit{et al.}~\cite{Belgacem:2018lbp} established a general framework for describing modified GW propagation through an effective GW luminosity distance and demonstrated how standard sirens can constrain such effects. Closely related to the present analysis, Lagos \textit{et al.}~\cite{Lagos:2019kds} investigated standard sirens in theories with a running Planck mass, where the resulting amplitude modification is formally similar to the rescaling considered here. Mastrogiovanni \textit{et al.}~\cite{Mastrogiovanni:2020mvm} subsequently placed observational constraints on anomalous GW friction using GW170817 and GW190521, explicitly discussing interpretations in terms of a time-varying Planck mass or Newton's constant. More broadly, Refs.~\cite{Ezquiaga:2018btd,Creminelli:2017sry,Ezquiaga:2017ekz} highlighted the profound impact of GW170817 on dark-energy and modified-gravity scenarios, emphasizing the role of multimessenger observations as probes of non-standard GW propagation. These studies provide an important context for the present work and motivate a careful comparison between propagation-dominated constraints and those arising from strong-field SEP tests.
}

In this work, we exploit multimessenger observations of BNS coalescences to test the strong equivalence principle through a possible temporal variation of the gravitational constant. A slowly varying $G$ provides a well-motivated and generic parameterization of SEP violation, with observable consequences for both the generation of GWs at the source and their propagation across cosmological distances. 
At the source, a time-dependent $G$ modifies the dynamics of compact binaries through its correction to gravitational self-energy, leading to characteristic imprints on the inspiral phase evolution. These effects can be described in terms of object-dependent sensitivities~\cite{Nordtvedt:1990zz}, which quantify the response of compact objects to variations in $G$. At the same time, a varying gravitational constant alters the propagation of GWs in an expanding Friedmann–Lemaître–Robertson–Walker (FLRW) universe, inducing additional modifications to the observed waveform amplitude. A consistent test of the SEP therefore requires a framework that simultaneously accounts for both source and propagation effects.

Building on this framework, we implement a physically consistent GW waveform model that incorporates both source and propagation effects of a slowly varying gravitational constant, while enforcing equation-of-state (EOS) dependent mass bounds through compact-object sensitivities. We apply this framework to GW170817 and perform a joint Bayesian analysis that combines GW data with independent electromagnetic constraints from the GRB 170817A counterpart. {By comparing source-only and propagation-only analyses, we find that both contributions constrain the varying-$G$ parameter at a comparable order of magnitude, while the final combined bound is mainly driven by the propagation-induced amplitude modification once the electromagnetic inclination prior is imposed.} We find no evidence for a temporal variation of the gravitational constant and constrain its fractional time derivative to $\dot{G}/G \in [-3.36 \times 10^{-9}, 5.34\times10^{-10}]~{\rm yr^{-1}}$. These constraints, derived from real multimessenger observations {and from a waveform model consistently including both source and propagation effects, provide a controlled test of the SEP with compact binary mergers.}

\begin{figure}[h]
\centering
\includegraphics[width=0.9\textwidth]{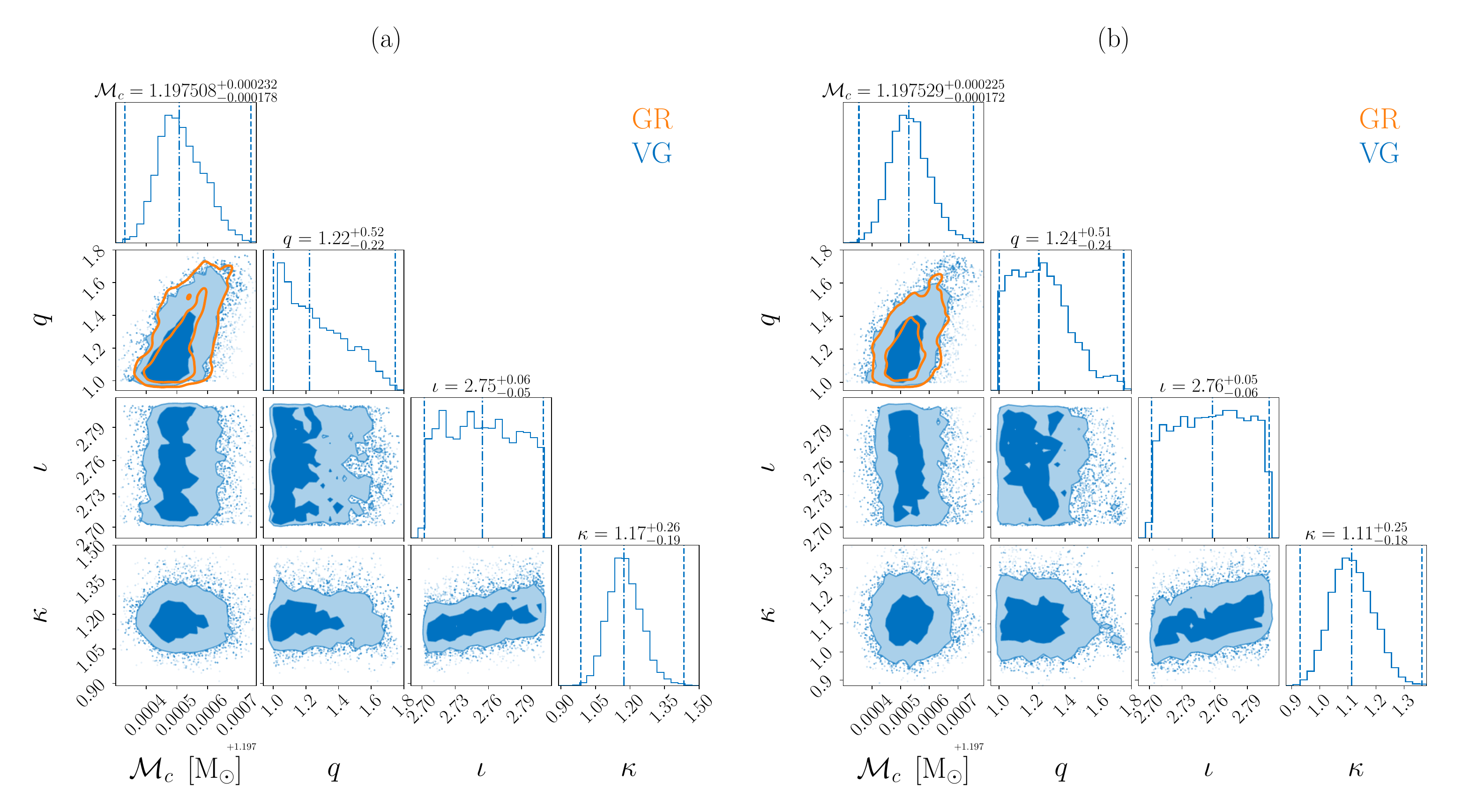}
\caption{\textbf{Posterior distributions from multimessenger constraints on a varying gravitational constant.} Posterior distributions of the key parameters are inferred from GW170817 using the \texttt{TaylorF2} (panel (a)) and \texttt{IMRPhenomXAS\_NRTidalv2} (panel (b)) waveform models. 
    {The two-dimensional contours show the 50\% and 90\% credible regions, while in the one-dimensional marginalized posteriors the dashed vertical lines denote the {99.7\%} interval and the central dash-dotted vertical line marks the median value.}
    Results obtained under GR are shown in orange, and those obtained within the varying-$G$ framework are shown in blue.}
    \label{results}
\end{figure}

\section*{Results and discussion}

\reply{Previous studies have explored several ways of using GW observations to constrain a varying gravitational constant. In the work of Ref.~\cite{Tahura:2019dgr}, $\dot{G}/G$ was constrained directly by incorporating varying-$G$ corrections into the gravitational-wave phase evolution. Ref.~\cite{Vijaykumar:2020nzc} constrained the source-to-observer gravitational constant ratio through the varying-$G$ modification of neutron-star masses in GW170817, and projected it into a bound on $\dot{G}/G$ using the lookback time. Refs.~\cite{An:2023rqz, Sun:2023bvy, An:2025cmm} instead considered varying-$G$ corrections to GW propagation and derived projected bounds using the luminosity-distance posterior of GW170817.   In this work, we jointly incorporate the varying-$G$ effects on neutron-star masses, GW phase, and GW propagation in an FLRW universe, and perform a full Bayesian analysis of the multimessenger observations of GW170817 and GRB~170817A to directly constrain the source-to-observer gravitational-coupling ratio. Focusing on the tighter one-sided bound on $\dot{G}/G$, namely the upper bound, our result is tighter by approximately fourteen, two, and one orders of magnitude than the three classes of previous results discussed above, respectively.  }

We analyze GW170817 strain data from LIGO Hanford and Livingston detectors, obtained from the Gravitational Wave Open Science Center (GWOSC)~\cite{GWOSC_GWTC}. The data are processed with a low-frequency cutoff of 20~Hz and a reference frequency of 40~Hz. The reference frequency is only a waveform-generation convention and does not enter the treatment of the observed data. For consistency, the same reference frequency is used for both the GR and varying-\(G\) waveforms. Bayesian parameter inference is performed using the \texttt{PyCBC} framework~\cite{Biwer:2018osg}. {Since the LIGO-Virgo Collaboration analysis of GW170817 found no strong evidence for spin precession and showed that precessing and non-precessing waveform models yield mutually consistent posterior distributions for the source parameters relevant to this work~\cite{LIGOScientific:2018hze}, we model GW170817 with non-precessing tidal waveform approximants in the present analysis for computational efficiency.
} 
{To ensure the robustness of our results, we employ two baseline GR waveform models, \texttt{TaylorF2}~\cite{Droz:1999qx, Buonanno:2009zt} and \texttt{IMRPhenomXAS\_NRTidalv2}~\cite{Colleoni:2023ple}. Although the observed signal is dominated by the inspiral phase, merger–ringdown effects are also incorporated through a continuous ($C^0$) extension within \texttt{IMRPhenomXAS\_NRTidalv2}, in analogy with the technique used in recent studies to handle numerical waveforms~\cite{Bonilla:2022dyt,Song:2025dbf}.}
In the Bayesian analysis, we incorporate independent electromagnetic constraints on the sky localization, luminosity distance, and orbital inclination of GW170817~\cite{DES:2017kbs, Cantiello:2018ffy, Mooley:2017enz, Mooley:2018qfh, Mooley:2022uqa}, enabling a genuine multimessenger inference that breaks key degeneracies inherent to GW–only analysis.

The posterior distributions of the key parameters are shown in Fig.~\ref{results}. The left and right panels present the results obtained with the \texttt{TaylorF2} and \texttt{IMRPhenomXAS\_NRTidalv2} waveforms, respectively. In both cases, the GR posteriors for the chirp mass $\mathcal{M}_c$ and mass ratio $q$ are highlighted in orange. The inferred values of $\mathcal{M}_c$ and $q$ are consistent across waveform models and agree with previously reported results~\cite{Nitz:2021zwj}. 

The corresponding posteriors obtained within the varying-$G$ framework are shown in blue. For both waveform families, the inferred values of $\mathcal{M}_c$, $q$, and the remaining intrinsic and extrinsic parameters remain fully consistent with their GR counterparts. No statistically significant shift is observed between the GR and varying-$G$ analyses, nor between the two waveform models, indicating that the posterior distributions are stable against both modeling choices and the inclusion of a varying gravitational constant. {We emphasize that this comparison is not intended as a symmetric model comparison under identical priors, but rather as a baseline consistency check. The agreement of the GR posteriors with previous LIGO--Virgo results validates our inference pipeline, while the consistency between the GR and varying-$G$ intrinsic parameters shows that the varying-$G$ corrections remain perturbatively small and do not significantly alter the binary properties, as assumed in the ppE framework.}

The non-GR parameter $\kappa = G_*/G_0$, which quantifies the ratio of the gravitational constant at the source to that measured at the detector, is constrained to $\kappa = 1.17^{+0.27}_{-0.19}$ and $\kappa = 1.11^{+0.25}_{-0.18}$ ({99.7\%} credible intervals) using the \texttt{TaylorF2} and \texttt{IMRPhenomXAS\_NRTidalv2} waveforms, respectively. The close agreement between these independent waveform models further supports the robustness of the inferred constraints. Using the lookback time of GW170817, these results translate into bounds on the temporal variation of the gravitational constant of $[-3.36 \times 10^{-9}, 1.53\times10^{-10}]~{\rm yr^{-1}}$ and $[-2.75 \times 10^{-9}, 5.34\times10^{-10}]~{\rm yr^{-1}}$, respectively. Adopting a conservative interpretation, we obtain a 
{conservative envelope of the two waveform-specific}
of $\dot{G}/G$ of $[-3.36 \times 10^{-9}, 5.34\times10^{-10}]~{\rm yr^{-1}}$. 

It is instructive to compare our results with previous GW-based studies that constrain a possible temporal variation of the gravitational constant~\cite{Chamberlain:2017fjl, Tahura:2018zuq, Tahura:2019dgr,Vijaykumar:2020nzc,  Barbieri:2022zge,Wang:2022yxb, An:2023rqz, Sun:2023bvy, Yuan:2024duo, An:2025cmm}. Early analyses based on Fisher-matrix techniques and Monte Carlo simulations for GW151226, which incorporated both amplitude and phase corrections within the parametrized post-Einsteinian framework, yielded comparatively weak bounds of $|\dot{G}/G| \lesssim 2.20\times10^{4}~{\rm yr^{-1}}$~\cite{Tahura:2019dgr}. Subsequent work constrained $\dot{G}/G$ by imposing physically allowed neutron-star mass ranges, derived from the Tolman–Oppenheimer–Volkoff (TOV) equations, on the projected GR mass posteriors of GW170817, obtaining bounds at the level of $\dot{G}/G \sim [-7 \times 10^{-9}, 5\times10^{-8}]~{\rm yr^{-1}}$~\cite{Vijaykumar:2020nzc}. Other studies exploited varying-$G$ effects on 
GW propagation, using projected GR luminosity distance posteriors of GW170817 to infer constraints of order $|\dot{G}/G| \lesssim 10^{-9}~{\rm yr^{-1}}$~\cite{An:2023rqz, Sun:2023bvy, An:2025cmm}. By contrast, our analysis performs a full Bayesian inference using the observed GW170817 data, incorporating independent electromagnetic constraints on the source geometry and distance, while simultaneously accounting for both source-generation and propagation effects of a varying gravitational constant and enforcing neutron-star mass bounds derived from TOV equations. This approach improves upon earlier Fisher-matrix estimates and confirms, with real multimessenger data, the projected constraints inferred from GR posteriors of GW170817.

\reply{The Fig.~\ref{results} shows an apparent preference for $\kappa>1$, which is mainly induced by the $\kappa$--$\iota$ degeneracy in the waveform amplitude. In this work, we have used the independent electromagnetic constraint on the inclination from the GRB~170817A afterglow to mitigate this degeneracy~\cite{Mooley:2017enz,Mooley:2018qfh,Mooley:2022uqa}. This constraint from GRB is narrower than the GR-based GW posterior and overlaps mainly with its upper part, so the residual $\kappa$-$\iota$ correlation shifts the marginalized $\kappa$ posterior slightly above one. Nevertheless, the Bayes factor still favors GR over the varying-$G$ model.}
{Meanwhile, it is worth emphasizing that the main result presented above is obtained using the \texttt{H4} EOS. In principle, the neutron-star sensitivity depends on the choice of EOS. Therefore, based on the \texttt{TaylorF2} waveform, we have repeated the analysis with several additional representative EOSs suggested in the detailed LIGO-Virgo analysis of GW170817~\cite{LIGOScientific:2018cki}, including \texttt{WFF1}, \texttt{APR4}, \texttt{SLy4}, and \texttt{MPA1}. The results obtained with these four additional EOSs are fully consistent with the \texttt{H4}-based results shown in Fig.~\ref{results} for all parameters. In particular, the central value of $\kappa$ differs by only about $1\%$. This demonstrates that our constraint is robust against the choice of EOS.}

In summary, we present a comprehensive multimessenger analysis that constrains a possible temporal variation of the gravitational constant as a test of the SEP.
We find no evidence for a temporal variation of $G$, and place a conservative {99.7\%} bound of $\dot{G}/G \in [-3.36 \times 10^{-9}, 5.34\times10^{-10}]~{\rm yr^{-1}}$, representing the most stringent constraints obtained to date from real GW observations.

\section*{Related physics in varying-$G$ theory}
\label{physics}

\subsection*{Modified GW waveform templates in varying-$G$ theory}
To model the impact of a time-varying gravitational constant on GW signals, we adopt the parametrized post-Einsteinian (ppE) framework~\cite{Yunes:2009ke,Tahura:2018zuq}.
In the frequency domain, the ppE waveform for a compact binary inspiral can be written as~\cite{Yunes:2009ke}
\begin{equation}
    \tilde{h}(f) = \tilde{h}_{\mathrm{GR}}(f) \left(1 + \alpha u^a \right) e^{i \beta u^b},
\end{equation}
where $\tilde{h}_{\mathrm{GR}}(f)$ denotes the GR waveform. The terms $\alpha u^a$ and $\delta\Psi = \beta u^b$ represent non-GR corrections to the waveform amplitude and phase, respectively. 
{Here \(u\equiv(\pi\mathcal{M}_cf)^{1/3}\), where  $f$ is the observed GW frequency and \(\mathcal{M}_c\equiv(m_{1}m_{2})^{3/5}/(m_{1}+m_{2})^{1/5}\) is the detector-frame chirp mass defined in terms of the detector-frame component masses $m_1$ and $m_2$.}
Deviations from GR are fully characterized by the ppE parameters $\{\alpha, \beta, a, b\}$. 
{
The possibility of constraining a time-varying Newton constant with GW observations was first developed by Yunes, Pretorius, and Spergel~\cite{Yunes:2009bv}. Building on this foundational work, the explicit ppE corrections induced by a varying gravitational constant, including the effects of compact-object sensitivities, were derived in Ref.~\cite{Tahura:2018zuq}; here we briefly summarize the key physical ingredients relevant for our analysis.
}

In varying-$G$ theories, the gravitational self-energy of a compact object depends explicitly on the gravitational constant. As a result, the mass of a body becomes time dependent when $G$ varies in time.
This induces an anomalous acceleration for a compact object moving with velocity $\mathbf{v}$ in the cosmic frame, which in turn affects the orbital angular momentum of a binary system~\cite{Nordtvedt:1990zz}.
The response of a compact object to a variation in $G$ is quantified by its sensitivity,
\begin{equation}
s=-\frac{G}{m}\frac{\delta m}{\delta G}=-\frac{\delta \ln m}{\delta\ln G},
\end{equation}
{where $m$ denotes the mass of the compact object.}

The anomalous acceleration $\mathbf{a}$ can be then written as~\cite{Nordtvedt:1990zz}
\begin{equation}
    \mathbf{a} = s\frac{\dot{G}}{G}\mathbf{v},
\end{equation}
where the overdot denotes a derivative with respect to time.
For a binary system with total mass denoted as $M=m_1+m_2$ and total angular momentum denoted as $j$, the time evolution of these quantities induced by a varying-$G$ is given by~\cite{Nordtvedt:1990zz}
\begin{align}
\dot{M}&=-\frac{m_{1}s_{1}+m_{2}s_{2}}{m_{1}+m_{2}}\frac{\dot{G}}{G}M,\\
\dot{j}&=\frac{m_{1}s_{1}+m_{2}s_{2}}{m_{1}+m_{2}}\frac{\dot{G}} {G}j,
\end{align}
where \(j=\sqrt{GMr}\) and $r$ denotes the binary separation. The combined variation of the total mass, gravitational constant, and angular momentum modifies the orbital separation at a rate
\begin{equation}
    \dot{r}_{\dot{G}}=-\left(\frac{\dot{G}}{G}+\frac{\dot{M}}{M}-2\frac{\dot{j}}{j}\right)r.
\end{equation}
The full evolution of the binary separation is therefore
\begin{equation}
    \dot{r} = \dot{r}_{\rm GW}+\dot{r}_{\dot{G}},\label{eq:rdot}
\end{equation}
where the contribution from GW emission is
\begin{equation}
    \dot{r}_{\mathrm{GW}}=-\frac{64}{5}\frac{G^3\mu M^2}{r^3},
\end{equation}
with $\mu=m_1m_2/M$ the reduced mass.
Using the modified orbital evolution in Eq.~(\ref{eq:rdot}), 
the ppE phase parameters associated with a varying gravitational constant are obtained as~\cite{Tahura:2018zuq}
\begin{align}
    b=&-13,\\
    \beta=&-\frac{25}{851968}\eta^{3/5}\frac{\dot{G}}{G}[(11+3(s_{1}+s_{2}))GM -41(Gm_{1}s_{1}+Gm_{2}s_{2})],
\end{align}
while the corresponding ppE amplitude parameters are~\cite{Tahura:2018zuq}
\begin{align}
    a= &-8, \\
    \alpha =& \frac{5}{512}\eta^{3/5}\frac{\dot{G}}{G}\left[(-7+s_{1}+s_{2})GM
    +13(Gm_{1}s_{1}+Gm_{2}s_{2})\right],
\end{align}
where \(\eta=m_1m_2/M^2\) is the symmetric mass ratio. In this framework, the parameter of primary interest is the fractional time variation of the gravitational constant, $\dot{G}/G$, which we constrain using GW observations.

\subsection*{GW propagating in varying-$G$ FLRW universe}

In many GW  tests of modified gravity, propagation effects are frequently overlooked.
However, if the gravitational coupling varies with cosmic time, the evolution of $G$ during GW propagation can lead to observable modifications to the waveform amplitude. In particular, Refs.~\cite{An:2023rqz, Sun:2023bvy} showed that, in varying-$G$ theories, the GW amplitude acquires a correction factor of \(\sqrt{G_0/G_*}\) {when propagating in a flat spacetime}, 
where \(G_*\)  and  \(G_0\) denote the gravitational constant at the source and at the detector, respectively. 
{Nevertheless, whether additional corrections emerge during propagation within a Friedmann-Lemaître-Robertson-Walker (FLRW) spacetime remains unexplored. In this section, we present a rigorous derivation to address this open question.}

{
Since we adopt a model-independent approach to test varying-$G$ theories, wherein the underlying dynamics driving the evolution of $G$ are disregarded, and the variation of $G$ is treated purely as an external input, it is of paramount importance to employ an appropriate framework to capture its physical impacts. In Ref.~\cite{Sun:2023bvy}, the authors parameterized these effects by linearizing the action and deriving the GW equations of motion under the assumption that $G$ depends solely on the time coordinate. In this work, we follow a similar yet more generalized heuristic: we vary the action assuming $G$ to be a general function of spacetime coordinates to obtain the modified field equations, and subsequently perform metric perturbations around an FLRW background to establish the resulting GW propagation equation.
}

{
We begin our derivation with the standard Einstein-Hilbert action, but with a spacetime-dependent gravitational coupling $G$ as
\begin{equation}
S=\int\mathrm{d}x^4 \sqrt{-g} \frac{R}{16\pi G}.
\end{equation}
Under a varying \(G\), the variation of the above action yields the equation of motion for the metric as
\begin{equation}
\begin{aligned}
R_{\mu\nu}-\frac{1}{2}g_{\mu\nu}R-\frac{2(\nabla_\mu  G)(\nabla_\nu G)}{G^2}+\frac{\nabla_\nu \nabla_\mu G}{G}-\frac{g_{\mu\nu}(\nabla_\rho \nabla^\rho G)}{G}+\frac{2g_{\mu\nu}(\nabla_\rho G)(\nabla^\rho G)}{G^2}=0.
\end{aligned}
\end{equation}
This equation is applicable not only to a varying-$G$ with explicit time dependence, but also to a completely general varying-$G$ that depends on the spacetime coordinates.
Performing the trace-reversal of the equation, we obtain
\begin{equation}
\begin{aligned}
R_{\mu\nu}-\frac{2\left(\nabla_{\mu}G\right)\left(\nabla_{\nu}G\right)}{G^{2}}+\frac{\nabla_{\nu}\nabla_{\mu}G}{G}+\frac{g_{\mu\nu}\left(\nabla_{\rho}\nabla^{\rho}G\right)}{2G}-\frac{g_{\mu\nu}\left(\nabla_{\rho}G\right)\left(\nabla^{\rho}G\right)}{G^{2}}=0. \label{eq:RmnG}
\end{aligned}
\end{equation}
Expanding the field equations (\ref{eq:RmnG}) to linear order in \(h_{\mu\nu}\), we obtain
\begin{equation}
\begin{aligned}
0=&- \tfrac{1}{2} \nabla_{\nu}\nabla_{\mu}h^{\rho }{}_{\rho } + \tfrac{1}{2} \nabla_{\rho }\nabla_{\mu}h^{}{}_{\nu}{}^{\rho } + \tfrac{1}{2} \nabla_{\rho }\nabla_{\nu}h^{}{}_{\mu}{}^{\rho } - \tfrac{1}{2} \nabla_{\rho }\nabla^{\rho }h^{}{}_{\mu\nu} + \frac{h^{}{}_{\mu\nu} \nabla_{\rho }\nabla^{\rho }G}{2 G}  - \frac{\nabla_{\mu}h^{}{}_{\nu \rho} \nabla^{\rho }G}{2 G} - \frac{\nabla_{\nu}h^{}{}_{\mu \rho} \nabla^{\rho }G}{2 G}\\
& + \frac{\nabla_{\rho }h^{}{}_{\mu\nu} \nabla^{\rho }G}{2 G} + \frac{g_{\mu\nu} \nabla_{\rho }h^{\sigma }{}_{\sigma } \nabla^{\rho }G}{4 G} - \frac{h^{}{}_{\mu\nu} \nabla_{\rho }G \nabla^{\rho }G}{G^2}  - \frac{g_{\mu\nu} \nabla^{\rho }G \nabla_{\sigma }h^{}{}_{\rho }{}^{\sigma }}{2 G}+ \frac{h^{}{}_{\rho \sigma } g_{\mu\nu} \nabla^{\rho }G \nabla^{\sigma }G}{G^2}
- \frac{h^{}{}_{\rho \sigma } g_{\mu\nu} \nabla^{\sigma }\nabla^{\rho }G}{2 G},
\end{aligned}\label{eq:pert}
\end{equation}
where the covariant derivative is defined with respect to the background metric $g_{\mu\nu}$.
}

We then consider the physical metric $\tilde{g}_{\mu\nu}$ as a perturbation around a homogeneous and isotropic FLRW background,
\begin{equation}
\tilde{g}_{\mu\nu}=g_{\mu\nu}+h_{\mu\nu},
\end{equation}
where \(g_{\mu\nu}=\rm{diag}(-a^2(\eta),a^2(\eta),a^2(\eta),a^2(\eta))\) is the background FLRW metric in the comoving coordinate \((\eta, x,y,z)\), $a(\eta)$ is the scale factor,
and \(h_{\mu\nu}\) denotes the the metric perturbation.
For simplicity, we consider a GW propagating along the $z$-direction and work in the transverse–traceless gauge. The nonvanishing components of the metric perturbation are
\begin{equation}
\begin{aligned}
h_{11}&=-h_{22}=a(\eta) h_{+}(\eta,z), \\
h_{12}&=h_{21}=a(\eta) h_{\times}(\eta,z),
\end{aligned}
\end{equation}
with other components to be zero.
Substitute this ansatz into into Eq.~(\ref{eq:pert}), we find that the $(1,1),~(1,2),~(2,1)~\mathrm{and}~(2,2)$ components yield identical equations for the two GW polarizations \(h_{+}\) and \(h_{\times}\),
\begin{equation}
\begin{aligned}
\partial_\eta^2 h_{+/\times}&+\left(\frac{2a^{\prime}}{a}-\frac{G^{\prime}}{G}\right)\partial_\eta h_{+/\times}-\partial_z^2 h_{+/\times}+\left(\frac{2a^{\prime\prime}}{a}+\frac{2a^{\prime 2}}{a^{2}}-\frac{4a^{\prime}G^{\prime}}{aG}-\frac{G^{\prime\prime}}{G}+\frac{2G^{\prime 2}}{G^{2}}\right)h_{+/\times}=0,
\end{aligned}
\end{equation}
where primes denote derivatives with respect to conformal time $\eta$.
Assuming a plane-wave solution \(h(\eta,z)=A(\eta)\mathrm{e}^{i k z}\), the amplitude $A(\eta)$ satisfies
\begin{equation}
\begin{aligned}
A''&+\left(\frac{2a^{\prime}}{a}-\frac{G^{\prime}}{G}\right)A'+\left(k^2+\frac{2a^{\prime\prime}}{a}+\frac{2a^{\prime 2}}{a^{2}}-\frac{4a^{\prime}G^{\prime}}{aG}-\frac{G^{\prime\prime}}{G}+\frac{2G^{\prime 2}}{G^{2}}\right)A=0.
\end{aligned}
\end{equation}
Introducing the redefinition \(A(\eta)=B(\eta)\frac{\sqrt{G(\eta)}}{a(\eta)}\), the above equation reduces to 
\begin{equation}
\begin{aligned}
B''+&\left( k^2 +\frac{a^{\prime\prime}}{a}+\frac{2a^{\prime 2}}{a^{2}}\right.\left.-\frac{3a^{\prime}G^{\prime}}{aG}-\frac{G^{\prime\prime}}{2G}+\frac{5G^{\prime 2}}{4G^{2}}\right)B=0.
\end{aligned}
\end{equation}
For GWs with wavelengths much shorter than the Hubble radius,
all terms in the brackets except $k^2$ are negligible, yielding \(B''+k^2B=0\). The resulting GW solution is therefore
\begin{equation}
h=\frac{\sqrt{G(\eta)}}{a(\eta)}\exp{\left(ik(z\pm\eta)\right)}.
\end{equation}

Denoting the GW amplitudes at the source and observer by \(A_{\rm s}\) and \(A_{\rm obs}\), we have the relation \(A_{\rm s} a(\eta_{\rm s})/\sqrt{G_*}=A_{\rm obs}a(\eta_o)/\sqrt{G_0}\), so we have
\begin{equation}
    A_{\rm obs} =\frac{a(\eta_{\rm s})}{a(\eta_o)
    }\sqrt{\frac{G_0}{G_*}}A_{\rm s}.
\end{equation}
The coefficient related to $a(\eta)$ corresponds to the standard cosmological redshift effect and is absorbed into the definition of the luminosity distance $D_\mathrm{L}$.
The additional factor \(\sqrt{G_0/G_*}\) represents a genuine propagation effect induced by the variation of the gravitational constant, in agreement with previous results obtained in Minkowski spacetime~\cite{An:2023rqz,Sun:2023bvy}.
{Therefore, similar to the flat spacetime scenario, the propagation of GWs within an FLRW background yields an identical modification factor to the waveform amplitude under varying-$G$ theories.}

{
Furthermore, given that our constraints on varying-$G$ theories rely on a multi-messenger framework, it is essential to investigate whether the variation of $G$ affects electromagnetic counterparts. In fact, within the current framework, such varying-$G$ effects have zero impact on optical observations.
Starting from the total action incorporating the electromagnetic (EM) field
\begin{equation}
S=\int\mathrm{d}x^4 \sqrt{-g} \left(\frac{R}{16\pi G}-\frac{1}{4}F_{\mu\nu}F^{\mu\nu}\right), \label{eq:GEM}
\end{equation}
where the varying gravitational coupling $G$ is assumed to enter only through the Einstein-Hilbert term and has no direct coupling to the EM sector, the resulting equations of motion for the EM field remain identical to those in GR as $\nabla_\mu F^{\mu\nu}=0$. 
By employing the geometric optics approximation, one can readily derive the photon trajectories from the field equation. Because the equations of motion are unaltered from GR, photons continue to propagate along standard null geodesics. Consequently, 
once the background metric is fixed, electromagnetic signal propagation depends solely on this metric and is not directly affected by the variation of $G$.
Admittedly, a time-dependent $G$ could in principle alter standard cosmological evolution. However, current observational constraints demonstrate that the observed cosmological metric aligns remarkably well with the standard $\Lambda$CDM framework within present observational precision. Even if varying-$G$ effects influence the large-scale matter distribution, 
their impact on electromagnetic observations would enter only indirectly through possible modifications of the background geometry or metric perturbations. In the present analysis, we adopt the observed cosmological background, which is effectively described by the standard $\Lambda$CDM metric.
Under this premise, 
varying-$G$ effects do not introduce any additional direct modification to electromagnetic propagation.
}

\subsection*{Sensitivities for neutron stars}
\label{NS_sensitivity}
To construct the GW waveform templates in a theory with a time-varying gravitational constant, it is necessary to evaluate the sensitivities of compact objects.
In previous GW-based studies of varying-$G$ theories, the sensitivities of neutron stars have often been estimated using the empirical relation from Refs~\cite{Lattimer:2010zz, Zhu:2018etc}. In this approach, the sensitivity is assumed to scale linearly with the neutron-star mass and is given by
\begin{equation}
    s=0.16\left(\frac{\mathbf{M}}{1.33~\mathbf{M}_\odot}\right),\label{eq:emp}
\end{equation}
where $\mathbf{M}$ denotes the neutron-star mass and $\mathbf{M}_\odot$ is the solar mass.
While this prescription provides a useful approximation, a more accurate determination of the sensitivity requires a self-consistent treatment based on the neutron-star EOS.
Below we outline a numerical procedure to compute neutron-star sensitivities by solving the TOV equations for a given EOS.

Because any realistic temporal variation of $G$ is expected to be small on timescales relevant to neutron-star structure, its direct contribution to the field equations can be neglected. The internal structure of neutron stars can therefore be described by the standard TOV equations, which read as follows
\begin{equation}
\begin{aligned}
\frac{dm(r)}{dr}=&4\pi r^2\rho(r),\\
\frac{d\bar{m}_b(r)}{dr}=&4\pi r^2\epsilon_b(r)\left(1-\frac{2 G m(r)}{r}\right)^{-1/2},\\
\frac{dP(r)}{dr}=&-\frac{Gm(r)\rho(r)}{r^2}\left(1+\frac{P(r)}{\rho(r)}\right)\left(1+\frac{4\pi r^3 P(r)}{m(r)}\right)\left(1-\frac{2 G m(r)}{r}\right)^{-1},
\end{aligned}\label{eq:TOV}
\end{equation}
where \(m(r)\) and \(\bar{m}_b(r)\)  denote the gravitational mass and the  baryon mass enclosed by a radius \(r\), respectively,
and \(P(r)\), \(\rho(r)\) and \(\epsilon_b(r)\) are the pressure, density and rest mass density at \(r\).
Given an EOS specifying the relations among \(P(r)\), \(\rho(r)\), and \(\epsilon_b(r)\), and the center density \(\rho(0)=\rho_c\), solving Eq.~(\ref{eq:TOV}) yields the neutron-star radius \(R\), gravitational mass \(\mathbf{M}=m(R)\), and baryon mass \(\bar{\mathbf{M}}=\bar{m}_b(R)\).
We assume the baryon number is conserved in the varying-$G$ scenario, such that the baryon mass of a neutron star remains unchanged under variations of $G$.

The $G$-dependence of Eq.~(\ref{eq:TOV}) can be removed by introducing the rescaled variables
\begin{equation}
\begin{aligned}
&m=G^{-\frac{3}{2}}m^*,\\
&\bar{m}_b=G^{-\frac{3}{2}}\bar{m}_b^*,\\
&r=G^{-\frac{1}{2}}r^*, \label{eq:map}
\end{aligned}
\end{equation}
in terms of which the TOV equations become
\begin{equation}
\begin{aligned}
\frac{dm^*}{dr^*}=&4\pi {r^*}^2\rho,\\
\frac{d\bar{m}^*_b}{dr^*}=&4\pi r^2\epsilon_b\left(1-\frac{2 m^*}{r^*}\right)^{-1/2},\\
\frac{dP}{dr^*}=&-\frac{m^*\rho}{{r^*}^2}\left(1+\frac{P}{\rho}\right)\left(1+\frac{4\pi {r^*}^3 P}{m^*}\right)\left(1-\frac{2 m^*}{r^*}\right)^{-1},
\end{aligned}\label{eq:TOVnew}
\end{equation}
where $m^*$, $\bar{m}_b^*$, $P$, $\rho$ and $\epsilon_b$ are functions of $r^*$.
For a given central density \(\rho_c\) and the EOS of the neutron star, solving Eq.~(\ref{eq:TOVnew}) yields the quantities \(\mathbf{M}^*=m^*(R^*)\) and \(\bar{\mathbf{M}}^*=\bar{m}^*_b(R^*)\), where \(R^*\) is the stellar radius in the rescaled coordinates.

Now we define a function $f$ which maps the value between the rescaled gravitational mass and baryon mass,
\begin{equation}
    f:  \mathbf{M}^* \mapsto \bar{\mathbf{M}}^*,
\end{equation}
and its inverse map is denoted as $g$
\begin{equation}
    g:  \bar{\mathbf{M}}^* \mapsto \mathbf{M}^*.
\end{equation}
For a neutron star with mass $\mathbf{M}$ and the gravitational constant $G$, then we have
\(\mathbf{M}^*=G^{3/2}\mathbf{M}\), and \(\bar{\mathbf{M}}^*=f(\mathbf{M}^*)=f(G^{3/2}\mathbf{M})\).
The baryon mass for this neutron star is 
\begin{equation}
\bar{\mathbf{M}}=G^{-3/2}\bar{\mathbf{M}}^*=G^{-3/2}f(G^{3/2}\mathbf{M}).
\end{equation}
When the gravitational constant changes to $G_1=G+\delta G$, 
the baryon mass keep fixed, and correspondingly, \(\bar{\mathbf{M}}^*_1=G_1^{3/2}G^{-3/2}f(G^{3/2}\mathbf{M})\).
Thus we have \(\mathbf{M}^*_1=g(G_1^{3/2}G^{-3/2}f(G^{3/2}\mathbf{M}))\).
So the mass of the neutron star changes to
\begin{equation}
\mathbf{M}_1=G_1^{-3/2}g\left(G_1^{3/2}G^{-3/2}f(G^{3/2}\mathbf{M})\right).
\end{equation}
Hence the sensitivity can be expressed as
\begin{equation}
\begin{aligned}
s=&-\frac{G}{\mathbf{M}}\lim_{\delta G\to 0}\frac{\mathbf{M}_1-\mathbf{M}}{\delta G}\\
=
&\frac{3}{2}\left(1-\frac{1}{G^{3/2}\mathbf{M}}f(G^{3/2}\mathbf{M})g'\left(f(G^{3/2}\mathbf{M})\right)\right),\label{eq:sensitivity}
\end{aligned}   
\end{equation}
where the prime denotes the derivative. 

\begin{figure}[h]
\centering
\includegraphics[width=0.7\textwidth]{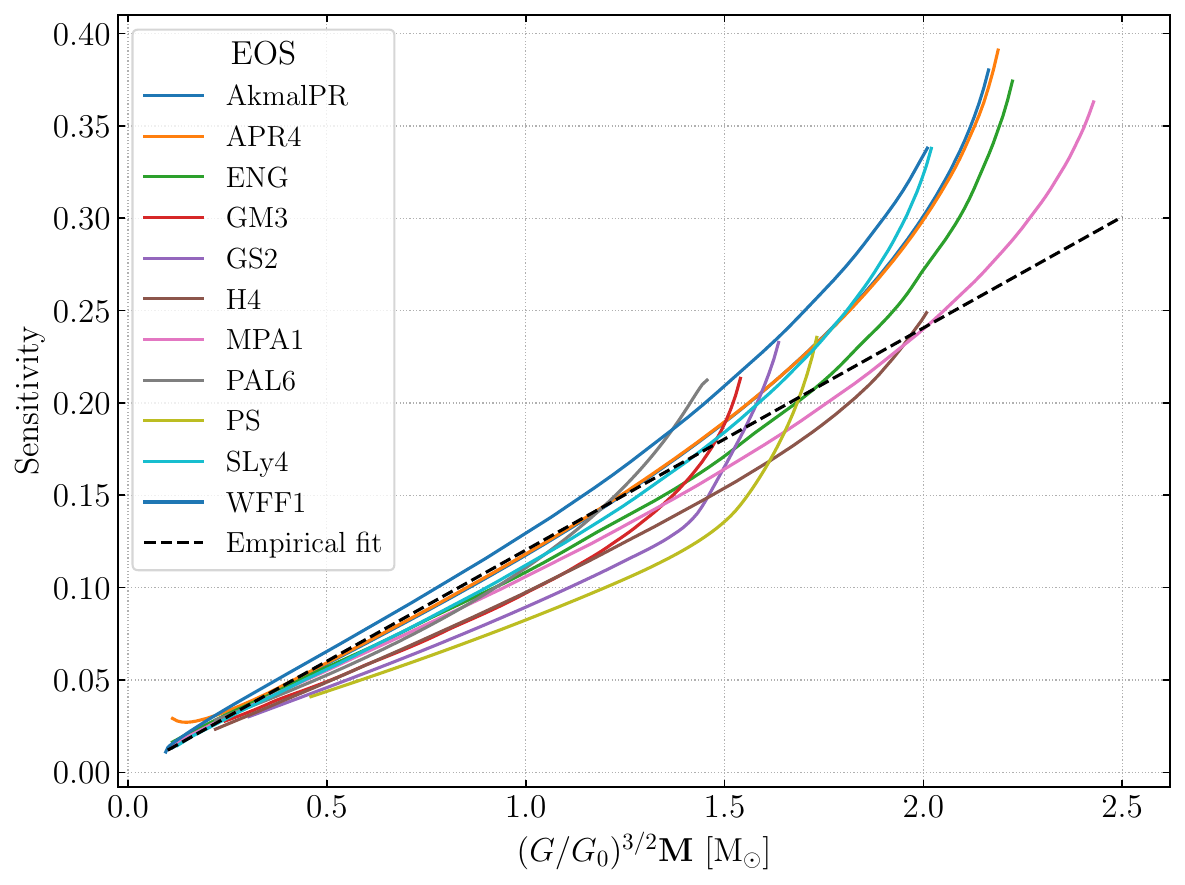}
\caption{\textbf{Neutron-star sensitivities as a function of {its} mass and gravitational constant for different equations of state.} Neutron-star sensitivities are computed for several representative equations of state. The black dashed line shows the empirical relation given in Ref.~\cite{Lattimer:2010zz, Zhu:2018etc}, while the colored curves correspond to sensitivities obtained from numerical solutions of the TOV equations. The EOSs are taken from Ref.~\cite{Lattimer:2000nx}.}
\label{sensitivities}
\end{figure}

We can see that for a given EOS, the sensitivity depends on \(G^{3/2}\mathbf{M}\).
To obtain the numerical results of the sensitivities, we first solve Eq~(\ref{eq:TOV}) with the current gravitational constant \(G_0\) and with different initial values \(\rho_c\) to obtain \(M(\rho_c)\) and \(\bar{M}(\rho_c)\). Then we perform interpolation on the different data points \((\mathbf{M}(\rho_c), \bar{\mathbf{M}}(\rho_c)) \) to obtain the function \(f\) and \(g'\), and finally we use Eq.~(\ref{eq:sensitivity}) with their variables \(\mathbf{M}\) replaced with \((G/G_0)^{3/2}\mathbf{M}\) to evaluate the sensitivities at different values of \(G\).
The resulting sensitivities for several representative EOSs are shown in Fig.~\ref{sensitivities}.
We find that the empirical relation in Eq.~(\ref{eq:emp}) provides a reasonable approximation at low masses, while for all EOSs considered, the sensitivity increases monotonically with mass but remains below $1/2$.

\subsection*{{Self-consistency checks via black hole sensitivities}}
\label{sensitivity_BH}

{
While the problem of black hole sensitivities falls outside the primary scope of our practical work, it serves as an ideal platform in this section to test the self-consistency of our established framework, wherein the underlying dynamics driving the evolution of $G$ are disregarded, and the variation of $G$ is treated purely as an external input.
A potential caveat of this model-independent framework is that, because the underlying dynamics driving the variation of $G$ (if any) are neglected, the variation of $G$ inevitably contributes an additional effective energy-momentum tensor upon varying the action. Crucially, there is no guarantee that this modified correction term satisfies the standard conservation law.
Therefore, this lack of conservation leads to an explicit contradiction with the geometric consistency required by the Bianchi identity.
A closely related parallel can be found in cosmological models featuring time-dependent dark energy; in the absence of an explicit dynamic framework for the dark energy sector, its energy-momentum tensor similarly violates the conventional conservation equations.
Despite the inherent non-conservation introduced by neglecting the full underlying dynamics, such a model-independent approach remains highly effective in capturing the leading-order physical effects. Crucially, the physical consistency of the theory is expected to be preserved at least up to the first-order correction of the varying-$G$ framework. As a concrete demonstration, the analysis of black hole sensitivities provides an ideal testbed to illustrate the validity and applicability of this method.
}

{Previous GW-based studies of varying-$G$ theories have conventionally adopted a canonical black-hole sensitivity value of $s=1/2$~\cite{Wang:2022yxb, Yuan:2024duo}. However, this specific value was originally established through highly model-dependent frameworks, primarily rooted in scalar–tensor theories. 
Specifically, within the Brans–Dicke formulation, {\it Will and Zaglauer}~\cite{Will:1989sk} argued that the sensitivity of a stationary black hole must equal $1/2$. 
The underlying heuristic relies on Hawking’s theorem~\cite{Hawking:1972qk}, which dictates that a stationary black hole cannot support nontrivial scalar hair. Consequently, the scalar field is forced to be constant in the black hole exterior, leading to a vanishing asymptotic scalar charge and thereby fixing $s=1/2$.
Remarkably, within our model-independent framework, we can replicate this exact result. 
In what follows, we provide a complementary and universally applicable argument demonstrating that the sensitivity of spherically symmetric black holes in varying-$G$ theories is identically $1/2$, thereby formally establishing the self-consistency of our phenomenological approach.}

{
Let us consider a slowly varying gravitational coupling $G(t)$, keeping only the leading-order terms in time derivatives. Assuming an adiabatic evolution, the spherically symmetric vacuum solution at each individual time slice can be matched to a Schwarzschild black hole, allowing the metric to be written as
}
\begin{equation}
g_{\mu\nu}=\mathrm{diag}\left(-f(r,t),f(r,t)^{-1}, r^2,r^2 \sin(\theta)^2\right),
\end{equation}
where \(f(r,t)=1-R_s(t)/r\) and \(R_s(t)=2G(t)M(t)\).
Substituting the metric into Eq.~(\ref{eq:RmnG}), the 
{$tr$-component}  
of the field equations yields
\begin{equation}
R_s(t)\dot{G}(t)-2G(t)\dot{R}_s(t)=0,
\end{equation}
where the {overdot denotes} the derivative {with respect to time}. Then we have,
\begin{equation}
    R_s(t)\propto\sqrt{G(t)}.\label{eq:RsqrtG}
\end{equation}
It follows that the black-hole mass evolves as
\begin{equation}
M(t)\propto\frac{R_s(t)}{G(t)}=G(t)^{-\frac{1}{2}},
\end{equation}
and therefore the sensitivity is
\begin{equation}
s=-\frac{\partial \ln M}{\partial \ln G}=\frac{1}{2}.
\end{equation}

{As a rigorous self-consistency check, we substitute the solution Eq.~\eqref{eq:RsqrtG} back into the modified field equations. We find that although the remaining components of the equations of motion do not vanish identically, they are comprised strictly of linear combinations of $\ddot{G}(t)$ and $\dot{G}(t)^2$. Consequently, within the first-order (or leading-order) approximation, these higher-order components reduce precisely to zero, meaning that the benchmark result for the black hole sensitivity is fully captured by the first-order $tr$-component of the field equations. This explicitly demonstrates once again that the first-order approximation of our model-independent framework is robust enough to inherit and capture the leading-order phenomenological physics of varying-$G$ theories.}

{Furthermore, this model-independent framework can be readily extended to the case of charged black holes. By introducing the Reissner–Nordstr\"{o}m (RN) metric into the same leading-order perturbation scheme, we find that the black hole sensitivity remains unaltered at $1/2$.}
For an RN black hole, based on the action~\eqref{eq:GEM}, we adopt the metric function 
\begin{equation}
    f(r,t)=1-\frac{R_m(t)}{r}+\frac{4\pi G(t)Q(t)^2}{r^2},
\end{equation}
together with the electromagnetic four-potential
\begin{equation}
    A_\mu=\left(\frac{Q(t)}{r},0,0,0\right).
\end{equation}
Assuming a slowly varying gravitational constant, we expand {the time-dependent parameters to the leading orders as}
\begin{equation}
\begin{aligned}
G(t)&=G_0(1+\lambda t)+O(t)^2,\\
R_m(t)&=R_0(1+\alpha \lambda t)+O(t)^2,\\
Q(t)&=Q_0(1+\beta\lambda t)+O(t)^2.
\end{aligned}
\end{equation}
{Upon substituting these expansions into the modified field equations and truncating at the leading order, the $r$-component of the Maxwell equations dictates $\beta=0$. This result implies that the black hole charge remains decoupled from the temporal variation of $G$ under the current framework, which aligns perfectly with our expectations.
In addition, the equations of motion for the metric enforce $\alpha=1/2$. Consequently, this implies that the model-independent framework universally yields a black-hole sensitivity of $1/2$, even in the presence of a net electric charge.}

\begin{figure}[h]
\centering
\includegraphics[width=0.68\textwidth]{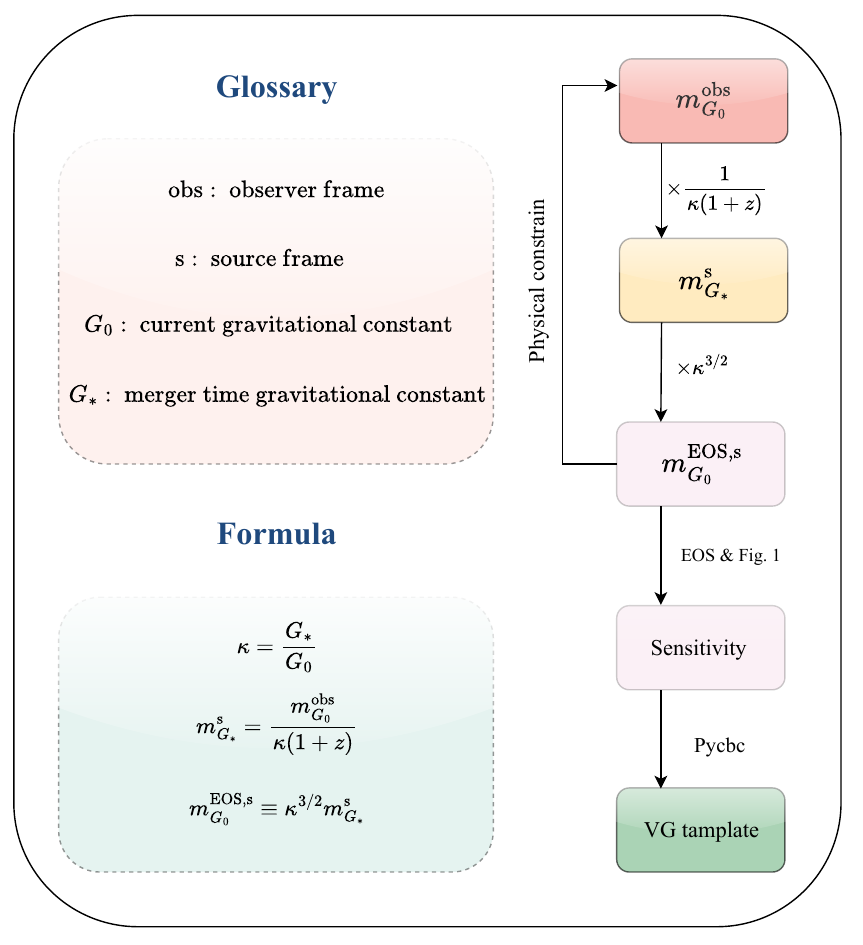}
\caption{\textbf{Workflow for constructing GW waveform templates in a varying-$G$ theory}. The varying-$G$ effects on neutron-star masses, GW phase evolution, and propagation are jointly incorporated into the waveform template used for Bayesian inference.}
\label{flow_chart}
\end{figure}

\section*{Methods}
In this section, we provide a detailed description of the varying-$G$ waveform template construction procedures and the corresponding Bayesian inference settings.

\subsection*{Varying-$G$ waveform template}
Under the assumption that the gravitational constant varies slowly on cosmological timescales, its time dependence can be treated as linear at leading order~\cite{Nordtvedt:1990zz, Tahura:2018zuq}.
The variation of the gravitational constant between the source and the detector can therefore be parametrized by
\begin{equation}
\kappa \equiv \frac{G_*}{G_0},
\end{equation}
denote the values of the gravitational constant at the source and at the detector, respectively. 
{In this work, we assume that the gravitational constant $G$ is slowly varying. Under this assumption, the fractional rate of change $\dot{G}/G$ can be linearly approximated as}
\begin{equation}
\frac{\dot{G}}{G}=\frac{G_0-G_*}{G_0 T}=\frac{1-\kappa}{T},
\end{equation}
where $T$ is the lookback time from the merger epoch to the present.
In our numerical implementation and Bayesian inference, we adopt $\kappa$ as the fundamental parameter characterizing varying-$G$ effects, and convert it to $\dot{G}/G$ when presenting the final constraints.

In the GR scenario, the GW waveforms depend on the dimensionless combination $GMf/c^3$, and the \texttt{LALSimulation}~\cite{lalsuite,swiglal} adopts natural units with $G=c=1$. 
Because of cosmological redshift, GW data analysis distinguishes between the source frame and the detector frame, related by
\begin{align}
    M_\mathrm{s} &= \frac{M_{\rm obs}}{1+z},\\
    f_\mathrm{s} &= f_{\rm obs}(1+z),
\end{align}
where $z$ is the redshift and subscripts ``s" and ``obs" denote the source and detector frames, respectively.

In the varying-$G$ scenario, 
analogous to the distinction between the source frame and the detector frame in GR, we define corresponding frames.
Denoting the source-frame parameters by \(\{G_*, m_{G_*}^{\rm s}, f_{\rm s}\}\) and the detector-frame parameters by \(\{G_0, m_{G_0}^{\rm obs}, f_{\rm obs}\}\), the invariance of the dimensionless combination implies
\begin{align}
    G_0 m_{G_0}^{\rm obs}f_{\rm obs}/c^3&=G_* m_{G_*}^{\rm s} f_{\rm s}/c^3,\\
    f_{\rm obs}&=\frac{f_{\rm s}}{1+z},
\end{align}
which yields
\begin{equation}
    m_{G_*}^{\rm s} = \frac{G_0}{G_*}\frac{m_{G_0}^{\rm obs}}{1+z}=\frac{m_{G_0}^{\rm obs}}{\kappa (1+z)}.
\end{equation}
Here $m_{G_0}^{\rm obs}$ is an auxiliary mass parameter, defined in the detector frame for numerical convenience, and serves as an output of the program with no direct physical meaning. Conversely, $m_{G_*}^{\rm s}$ is the true physical mass of the compact object at the source.

As shown in Sec.~\ref{NS_sensitivity}, the sensitivity of a neutron star depends on the combination \((G_*/G_0)^{3/2} m_{G_*}^{\rm s}=\kappa^{3/2}m_{G_*}^{\rm s}\).  We therefore define the corresponding EOS mass under the present gravitational constant  $G_0$ as
\begin{equation}
    m_{G_0}^{\mathrm{EOS},{\rm s}}\equiv\kappa^{3/2}m_{G_*}^{\rm s}=\frac{\kappa^{1/2}}{1+z}m_{G_0}^{\rm obs}.
\end{equation}
Following Ref.~\cite{Vijaykumar:2020nzc}, this quantity must lie within the physically allowed mass range determined by the neutron-star EOS evaluated at $G_0$.
In this work, we adopt the widely used \texttt{H4} EOS, which imposes the constraint
\begin{equation}
m_{G_0}^{\mathrm{EOS},{\rm s}} \in (0.08, 2.01)~{\rm M}_\odot,
\end{equation}
as illustrated in Fig.~\ref{sensitivities}. 

When matching waveform templates to GW data, the phase evolution plays the dominant role in parameter estimation, while amplitude corrections are often subleading. 
{In our analysis, the varying-$G$ correction to the gravitational-wave amplitude has two distinct origins. The ppE amplitude correction, $\alpha u^a$, arises from the generation effect and is found to be of order $10^{-16}$ across the Monte Carlo samples, and is therefore negligible. In contrast, the propagation effect induced by a varying gravitational constant introduces an amplitude correction factor of $1/\sqrt{\kappa}$, which is significant and must be retained.
}

Combining the ppE phase correction with the propagation effect, the detector-frame GW waveform template in the varying-$G$ scenario is given by
\begin{equation}
    \tilde{h}_{\rm VG}(f) = \frac{1}{\sqrt{\kappa}} \tilde{h}_{\mathrm{GR}} e^{i \beta_{\rm VG} u^{b_{\rm VG}}}, \label{ppewaveform}
\end{equation}
where $\tilde{h}_{\mathrm{GR}}$ is the GR waveform template generated by \texttt{PyCBC} using the detector-frame mass $m_{G_0}^{\rm obs}$,
and
\begin{align}
    b_{\rm VG}=&-13, \nonumber\\
    \beta_{\rm VG}=&-\frac{25}{851968}\eta^{3/5}\frac{1-\kappa}{T}G_*[(11+3(s_{1}+s_{2}))M_{G_*}^{\rm s} \nonumber-41(m_{G_*,1}^{\rm s} s_{1}+m_{G_*,2}^{\rm s} s_{2})]\nonumber\\
    =&-\frac{25}{851968}\eta^{3/5}\frac{1-\kappa}{T(1+z)}G_0[(11+3(s_{1}+s_{2}))M_{G_0}^{\rm obs} \nonumber-41(m_{G_0,1}^{\rm obs} s_{1}+m_{G_0,2}^{\rm obs} s_{2})],\label{ppE_phase}
\end{align}
with $s_1$ and $s_2$ evaluated using the EOS masses $m_{\rm G_0,1}^{\rm EOS,s}$ and $m_{\rm G_0,2}^{\rm EOS,s}$, respectively.
The overall workflow of the varying-$G$ waveform construction is summarized in Fig.~\ref{flow_chart}.

{As a final remark, we emphasize that the above waveform template relies on a slowly changing $G$, corresponding to a monotonic temporal evolution at the linear level. Moving beyond this, if we consider a $G$ that varies rapidly or fluctuates, the situation changes. However, because the modified waveform's core contributions stem from amplitude corrections and the neutron star equation of state, both of which depend directly on the explicit value of $\kappa$, our framework remains robust: if the variation of $G$ is non-monotonic, our bounds simply translate into a constraint on the time-averaged effect of such variations.}

\begin{table*}[ht]
\caption{\textbf{Prior distributions used in the varying-$G$ analysis of GW170817 with the \texttt{TaylorF2} and \texttt{IMRPhenomXAS\_NRTidalv2} waveform templates}.}
\label{priors}
\renewcommand{\arraystretch}{1.4}
\small
\begin{tabular}{lcccc}
\toprule
Parameters & \ \texttt{TaylorF2} (GR) \ & \ \texttt{TaylorF2} (VG) \ & \ \texttt{IMRPhenomXAS\_NRTidalv2} (GR) \ & \ \texttt{IMRPhenomXAS\_NRTidalv2} (VG) \ \\
\midrule
$\mathcal{M}_c~{[\rm M_{\odot}]}$& $\mathcal{U}[1.15, 1.23]$ & $\mathcal{U}[1.15, 1.23]$ & $\mathcal{U}[1.15, 1.23]$ & $\mathcal{U}[1.15, 1.23]$ \\
$q$& $\mathcal{U}[1.00, 2.00]$ & $\mathcal{U}[1.00, 2.00]$ & $\mathcal{U}[1.00, 2.00]$ & $\mathcal{U}[1.00, 2.00]$ \\
$\chi_{1z}$& $\mathcal{U}[-0.05, 0.05]$ & $\mathcal{U}[-0.05, 0.05]$ &  $\mathcal{U}[-0.05, 0.05]$ & $\mathcal{U}[-0.05, 0.05]$ \\
$\chi_{2z}$& $\mathcal{U}[-0.05, 0.05]$ & $\mathcal{U}[-0.05, 0.05]$ & $\mathcal{U}[-0.05, 0.05]$ & $\mathcal{U}[-0.05, 0.05]$ \\
$\Lambda_1$& $\mathcal{U}[0, 5000]$ & $\mathcal{U}[0, 5000]$ & $\mathcal{U}[0, 5000]$ & $\mathcal{U}[0, 5000]$ \\
$\Lambda_2$& $\mathcal{U}[0, 5000]$ & $\mathcal{U}[0, 5000]$ & $\mathcal{U}[0, 5000]$ & $\mathcal{U}[0, 5000]$ \\
$\kappa$& -- & ${\rm Log \ uniform}~[10^{-4}, 100]$ & -- & ${\rm Log \ uniform}~[10^{-4}, 100]$ \\
$D_L~{[\rm Mpc]}$& $40.7$ & $40.7$ & $40.7$ & $40.7$ \\
$\iota~{\rm [rad]}$& sine uniform &  $\mathcal{U}[2.70, 2.81]$  & sine uniform  &  $\mathcal{U}[2.70, 2.81]$\\
$\alpha~{\rm [rad]}$& 3.45 &3.45 & 3.45 & 3.45 \\
$\delta~{\rm [rad]}$& -0.41 & -0.41 & -0.41 & -0.41 \\
$\psi~{\rm [rad]}$& $\mathcal{U}[0, 2\pi]$ & $\mathcal{U}[0, 2\pi]$ & $\mathcal{U}[0, 2\pi]$ & $\mathcal{U}[0, 2\pi]$ \\
$\delta t_c~\mathrm{[s]}$& $\mathcal{U}[-0.1, 0.1]$ & $\mathcal{U}[-0.1, 0.1]$ &  $\mathcal{U}[-0.1, 0.1]$ &  $\mathcal{U}[-0.1, 0.1]$ \\
$\phi_c~{\rm [rad]}$& $\mathcal{U}[0, 2\pi]$ & $\mathcal{U}[0, 2\pi]$ & $\mathcal{U}[0, 2\pi]$ &  $\mathcal{U}[0, 2\pi]$\\
\bottomrule
\end{tabular}
\end{table*}

\subsection*{Bayesian inference}
In this work, we inference both GR and non-GR parameters from the observed GW data within a Bayesian framework~\cite{Finn:1992wt}. The posterior distribution of model parameters $\vartheta$, given the observed data $d$ and hypothesis $\mathcal{H}$ is,
\begin{equation}
p(\vartheta|d,\mathcal{H})=\frac{p(d|\vartheta,\mathcal{H})\mathrm{~}p(\vartheta|\mathcal{H})}{p(d|\mathcal{H})},
\end{equation}
where $p(d|\vartheta,\mathcal{H})$ is the likelihood, $p(\vartheta|\mathcal{H})$ the prior, and $p(d|\mathcal{H})$ the Bayesian evidence. 

For a network of multiple detectors, assuming stationary Gaussian noise, the log-likelihood takes the form
\begin{equation}
\log p(d|\vartheta,\mathcal{H})=\log\bar{\alpha}-\frac{1}{2}\sum_i\left\langle d_i-h_i(\vartheta)|d_i-h_i(\vartheta)\right\rangle,
\end{equation} 
where $\log\bar{\alpha}$ is the normalizing constant, the index $i$ labels individual detectors, and $h_i(\vartheta)$ denotes the waveform template projected onto the $i$-th detector. The inner product is defined as, 
\begin{equation}
    \langle a(t)|b(t)\rangle=2\int_{f_{\mathrm{low}}}^{f_{\mathrm{high}}}\frac{\tilde{a}^*(f)\tilde{b}(f)+\tilde{a}(f)\tilde{b}^*(f)}{S_n(f)}\mathrm{d}f.
\end{equation}
where $S_n(f)$ is the one-sided noise power spectral density, and $f_{\mathrm{low}}$ and $f_{\mathrm{high}}$ denote the lower and upper frequency cutoffs of the analysis, respectively.

{For the GW170817 analysis, we use the publicly available cleaned GWOSC strain data from the LIGO Hanford and Livingston detectors, H1 and L1~\cite{GWOSC_GWTC}. Virgo is not included in the parameter-estimation runs. The reported network SNR of GW170817 is \(\rho_{\rm net}=32.4\), with individual detector contributions \(\rho_{\rm H1}=18.8\), \(\rho_{\rm L1}=26.4\), and \(\rho_{\rm V1}=2.0\)~\cite{LIGOScientific:2017vwq}. Thus, Virgo contributes only \(2.0^2/32.4^2\simeq 0.4\%\) of the total \(\rho_{\rm net}^2\). Since the sky position is fixed independently by the electromagnetic counterpart, the low-SNR Virgo data do not provide significant additional information for the present parameter-estimation analysis, while including them would increase the computational cost. We therefore use only the H1 and L1 data. The Livingston data are taken in their glitch-subtracted form, with the short-duration glitch near the signal removed using a time-frequency wavelet reconstruction~\cite{Cornish:2014kda}. The likelihood is evaluated using a strain segment extending from \(512\,{\rm s}\) before to \(32\,{\rm s}\) after the merger time. The noise PSD is estimated from a longer stretch of data using the \texttt{PyCBC} \texttt{median-mean} method, with \texttt{psd-start-time = -1800}, \texttt{psd-end-time = 128}, and \texttt{psd-inverse-length = 8}.}

{
Our simulations indicate that the current varying-$G$ constraint is largely amplitude dominated, being primarily controlled by the propagation factor $1/\sqrt{\kappa}$. The parameter $\kappa \equiv G_\ast/G_0$ also enters through the leading $-4~{\rm PN}$ phase correction and through the EOS-dependent neutron-star structure quantities; these effects are included in the analysis, but at the current sensitivity they provide subdominant information compared with the propagation-amplitude contribution. Following the GW170817 calibration estimates and the improvement obtained when electromagnetic distance/inclination information is included~\cite{Essick_2019}, we assign a conservative residual amplitude-calibration uncertainty of order $15\%\text{--}20\%$ for the H1--L1 analysis. Propagating this uncertainty gives a relative calibration systematic of order $15\%\text{--}20\%$ in the inferred $\dot G/G$ interval. 
For computational efficiency, we do not model this effect explicitly in the data analysis. Instead, we use the calibrated strain data released by the LIGO--Virgo Collaboration, effectively keeping the residual calibration factors fixed in the main parameter estimation runs.
}

We consider the spin aligned BNS systems including tidal interactions. In this case, the GR waveform is described by the following set of 13 parameters:
\begin{equation}
\vartheta_{\rm GR}=\{\mathcal{M}_c,q,\chi_{1z},\chi_{2z}, \Lambda_1, \Lambda_2, D_L,\iota, \alpha,\delta,\psi,t_c,\phi_c\},
\end{equation}
where $\mathcal{M}_c$ and $q$ are chirp mass and mass ratio,
$\chi_{1z}$ and $\chi_{2z}$ are the dimensionless spin components aligned with the orbital angular momentum, $\Lambda_1$ and $\Lambda_2$ are the tidal deformabilities, $D_L$ is the luminosity distance, $\alpha$ and $\delta$ are the right ascension and declination angles, $\iota$ and $\psi$ denote the inclination angle the polarization angle, and $\phi_c$ and $t_c$ are the coalescence phase and time, respectively. \reply{It is worth noting that, in principle, varying-$G$ effects can also enter through the tidal deformabilities $\Lambda_i$, which are sampled independently in the present analysis for GW170817. When the $\kappa$ dependence of $\Lambda_i$ induced by the modified neutron-star structure is taken into account, we find that it slightly modifies the shape of the one-dimensional posterior distribution of $\kappa$, while having a negligible impact on the resulting bound on $\kappa$.}

In the varying-$G$ framework, an additional non-GR parameter $\kappa$ is introduced, such that
\begin{equation}
\vartheta_{\rm VG} = \vartheta_{\rm GR} + \{\kappa\}.
\end{equation}

We apply this framework to GW170817, the first observed BNS inspiral and the first gravitational-wave event with confirmed electromagnetic counterparts. Follow-up electromagnetic observations provide precise and independent constraints on several extrinsic parameters of the source. To improve the efficiency and robustness of the Bayesian inference, we fix the luminosity distance to $D_L = 40.7$~Mpc , inferred from surface brightness fluctuation measurement with the Hubble Space Telescope \cite{Cantiello:2018ffy}, corresponding to a lookback time of $T=1.31 \times 10^{8}$~years. We further fix the sky location to $\alpha = 3.45$~rad and $\delta = -0.41$~rad based on observations with the Dark Energy Camera~\cite{DES:2017kbs}. As a result, 10 and 11 parameters remain to be sampled in the GR and varying-$G$, respectively. 

As shown in Eq.~(\ref{ppewaveform}), the propagation effect induced by a varying gravitational constant introduces a multiplicative correction to the waveform amplitude, giving rise to degeneracies between the non-GR parameter $\kappa$ and both the luminosity distance $D_L$ and the inclination angle $\iota$. In this analysis, the degeneracy between $\kappa$ and $D_L$ is removed by fixing $D_L$ using independent electromagnetic distance measurements. A remaining degeneracy between $\kappa$ and $\iota$ therefore needs to be broken. To this end, we incorporate independent constraints on the inclination obtained from very long baseline interferometry (VLBI) observations of the 
GRB 170817A afterglow. These measurements constrain the inclination to the 90\% credible interval [2.70, 2.81]~rad~\cite{Mooley:2017enz, Mooley:2018qfh, Mooley:2022uqa}. We adopt this interval as the prior on $\iota$ in the varying-$G$ analysis, while a sine-uniform prior is used in the GR case. The complete prior distributions employed in both analyses are summarized in Table~\ref{priors}.

\begin{figure}[h]
\centering
\includegraphics[width=0.95\textwidth]{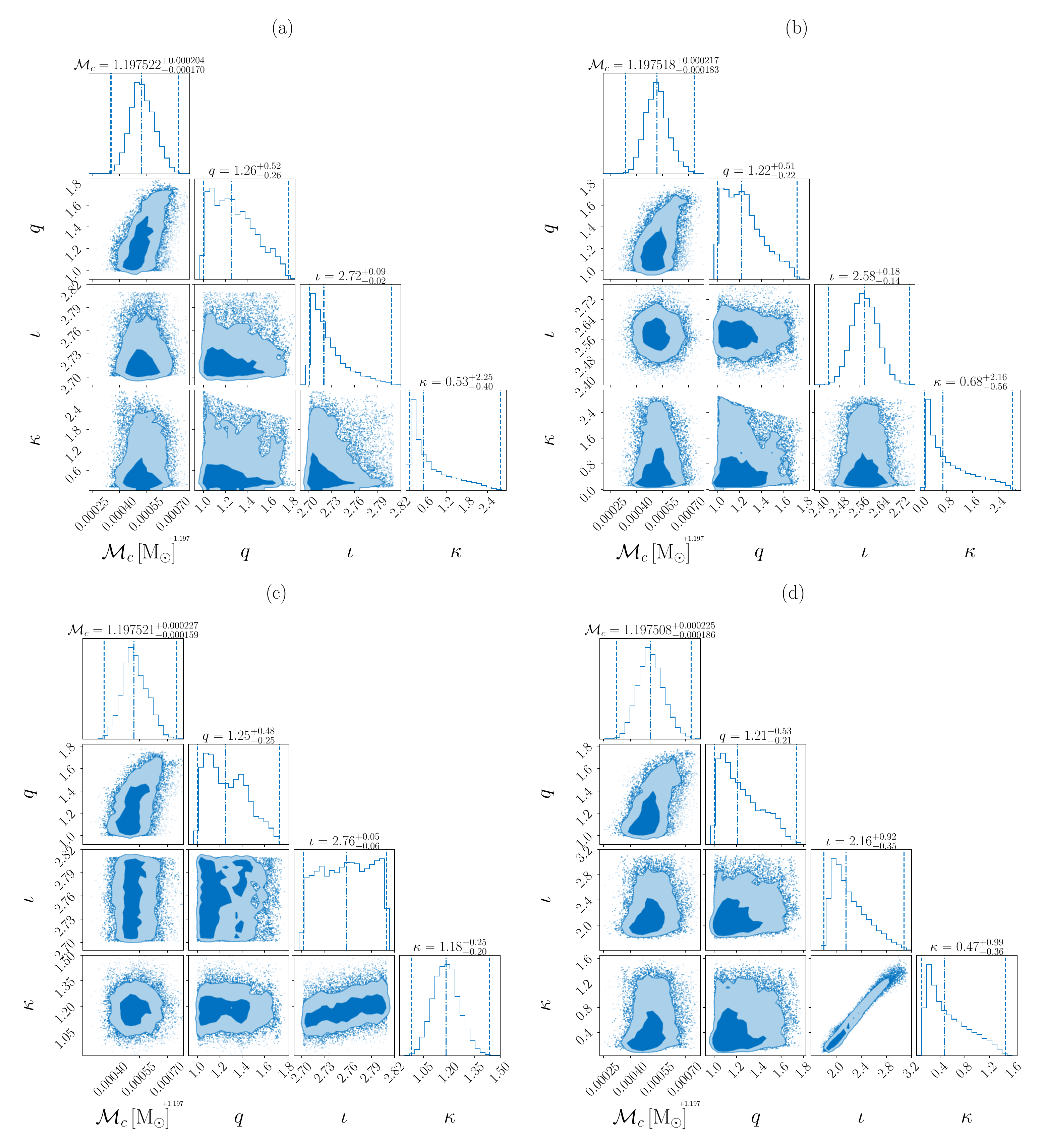}
\caption{{\textbf{Comparison of source-side and propagation-induced varying-$G$ constraints with different inclination priors.}
Panels (a) and (b) include only the source-side strong-field corrections with the EM and sine-uniform inclination priors, respectively, while panels (c) and (d) include only the propagation-induced amplitude correction with the same two inclination priors.}}
\label{prop_src}
\end{figure}

{
\subsection*{Degeneracy Analysis}
In the varying-$G$ framework considered in this work, the deviation from GR enters the GW analysis through two physically distinct channels. The first channel is the source-side strong-field dynamical effect. This includes the ppE correction induced by the time variation of $G$ in the binary evolution, which appears mainly in the waveform phase, and the varying-$G$ modification of the neutron-star EOS mass $m_{G_0}^{\mathrm{EOS},{\rm s}}$, which imposes an additional physical constraint on the sampled masses through the EOS-dependent allowed mass range. The second channel is the propagation effect, which modifies the observed GW amplitude through the factor $1/\sqrt{\kappa}$, as shown in Eq.~(\ref{ppewaveform}).}

{
To disentangle these effects and to clarify the role of the inclination prior, we performed additional MCMC analyses based on the \texttt{TaylorF2} waveform for the following four cases:
\begin{itemize}
\item[(a)] \textbf{Varying-G\_Src with EM $\iota$ prior}, in which only the two source-side strong-field corrections are included. The priors are the same as those used in the \texttt{TaylorF2} varying-$G$ case in Table~\ref{priors}. In particular, the inclination prior $\iota$ is taken from electromagnetic counterpart observations.
\item[(b)] \textbf{Varying-G\_Src with sine-uniform $\iota$ prior}, in which only the two source-side strong-field corrections are included. The priors are the same as in Table~\ref{priors}, except that the inclination prior $\iota$ is chosen to be sine-uniform.
\item[(c)] \textbf{Varying-G\_Prop with EM $\iota$ prior}, in which only the propagation effect is included. The priors are the same as in case (a).
\item[(d)] \textbf{Varying-G\_Prop with sine-uniform $\iota$ prior}, in which only the propagation effect is included. The priors are the same as in case (b).
\end{itemize}
}

{
The results are shown in Fig.~\ref{prop_src}. When only the two source-side strong-field effects are included, using the same prior choices as in the baseline varying-$G$ analysis, we obtain the constraint $\kappa = 0.53^{+2.36}_{-0.40}$, as shown in panel (a). This constraint is of the same order of magnitude as that obtained from the combined-effects analysis, $\kappa = 1.17^{+0.27}_{-0.19}$, although it is considerably looser. However, when only the source-side effects are considered, the strong degeneracy between $\kappa$ and $\iota$ induced by the propagation amplitude correction is absent. Therefore, in this case, a more natural and less restrictive choice is to adopt a generic sine-uniform prior on $\iota$. This case is shown in panel (b), where we obtain $\kappa = 0.68^{+2.15}_{-0.56}$. This result is again of the same order of magnitude as the combined-effects constraint.
}

{
A closer comparison between panels (a) and (b) shows that the posterior distributions of $\iota$ differ between the two cases, although the lower bound in panel (a) still overlaps with the upper bound in panel (b). This is not unexpected. The electromagnetic prior on $\iota$ adopted in the baseline varying-$G$ analysis is taken from independent constraints on the inclination obtained from very long baseline interferometry (VLBI) observations of the GRB~170817A afterglow. The electromagnetic analyses themselves~\cite{Mooley:2017enz, Mooley:2018qfh, Mooley:2022uqa} also noted that the inclination posterior inferred from electromagnetic observations is not identical to that obtained by the LIGO--Virgo Collaboration through GR waveform inference~\cite{LIGOScientific:2018hze}. This difference arises because the two analyses constrain related but not exactly identical quantities, adopt different priors, and because the GW-only inclination inference is affected by the well-known distance--inclination degeneracy. Nevertheless, since the upper bound of the GW-only constraint and the lower bound of the EM-only constraint still overlap, the electromagnetic inclination posterior can be regarded as a physically well-motivated independent observational input. Therefore, the difference between the $\iota$ posteriors in panels (a) and (b) can be attributed to the different choices of the inclination prior. If one is interested in isolating the source-side dynamical effects alone, the more appropriate result is the one shown in panel (b).
}

{
We then consider the cases in which only the propagation effect is included. As shown in panel (c), when the same priors as in the baseline combined-effects analysis are adopted, the posterior distributions of all parameters are almost identical to those obtained in the combined-effects case shown in Fig.~\ref{results}. This apparent dominance of the propagation effect can be understood from two aspects: first, the choice of the inclination prior, and second, the strong degeneracy between $\kappa$ and $\iota$ induced by the propagation amplitude correction. This is clearly illustrated in panel (d). If we adopt the same sine-uniform inclination prior as in the pure-GR inference, the degeneracy between $\kappa$ and $\iota$ causes the inferred inclination posterior to become significantly shifted relative to both the pure-GR inference and the electromagnetic constraint. Therefore, in order to break this degeneracy, the baseline varying-$G$ analysis uses the independently measured electromagnetic inclination posterior as an external input.
}

{
In summary, the analyses with only the varying-$G$ source-side dynamical effects and with only the varying-$G$ propagation effect both constrain $\kappa$ at the same order of magnitude. Therefore, if one aims to include the full varying-$G$ physics, these effects should be combined consistently. Because the propagation effect induces a strong $\kappa$--$\iota$ degeneracy, an independent electromagnetic constraint on $\iota$ is needed to obtain the combined result reported in Fig.~\ref{results}. Thus, while the final constraint is led by the propagation-induced amplitude effect, it is obtained within a framework that consistently includes the source-side strong-field contributions.
}

\emph{\textbf{Acknowledgements}.---}
This work is supported in part by National Natural Science Foundation of China under Grant No.~12335006 and in part by the National Natural Science Foundation of China under Grant No.~12547101. ZL is supported by ``the Fundamental Research Funds for the Central Universities''.
It is also supported by the High-performance Computing Platform of Peking University.

\emph{\textbf{Author Contributions }.---}
{J.Z. contributed to the conception of the study, discussions, calculations for Fig. 2, and the writing and revision of the section ``Related physics in varying-$G$ theory”. H.S. contributed to the conception of the study, discussions, calculations for Figs. 1, 3, and 4, and the writing and revision of the Abstract, Introduction, Results and Discussion, and Bayesian inference of varying-$G$ theory sections and the coordination of the whole project. Z.L. contributed to the conception of the study, discussions, and revision of the Bayesian inference of varying-$G$ theory section. H.L. contributed to the conception of the study, discussions, and revision of the Introduction. P.J., J.W., and H.Y. contributed to the conception of the study and discussions. B.M. contributed to the conception of the study, discussions, and provided supervision.}

\emph{\textbf{Data availability}.---}
{The data generated and analyzed during this study are available from the corresponding author upon reasonable request.}

\emph{\textbf{Code Availability }.---}
{The code used in this study is available from the corresponding author upon request.}

\emph{\textbf{Competing interests }.---}
{The authors declare no competing interests.}

\appendix

\bibliography{scibib.bib}
\end{document}